\documentstyle[aps]{revtex}
\draft
\begin{document}
\title{Nucleon momentum distribution in deuteron and other nuclei
within the light-front dynamics method}
\author{A.N. Antonov, M.K. Gaidarov, M.V. Ivanov, and D.N. Kadrev}
\address{Institute of Nuclear Research and Nuclear Energy,
Bulgarian Academy of Sciences, Sofia 1784, Bulgaria}
\author{G.Z. Krumova}
\address{University of Rousse, Rousse 7000, Bulgaria}
\author{P.E. Hodgson}
\address{Subdepartment of Nuclear and Particle Physics,
University of Oxford, Oxford OX1-3RH, U.K.}
\author{H.V. von Geramb}
\address{Theoretische Kernphysik, Universit$\ddot{a}$t Hamburg,
Hamburg D-22761, Germany}
\maketitle

\begin{abstract}
The relativistic light-front dynamics (LFD) method has been shown
to give a correct description of the most recent data for the
deuteron monopole and quadrupole charge form factors obtained at
the Jefferson Laboratory for elastic electron-deuteron scattering
for six values of the squared momentum transfer between 0.66 and
1.7 (GeV/c)$^{2}$. The good agreement with the data is in contrast
with the results of the existing non-relativistic approaches.

In this work we firstly make a complementary test of the LFD
applying it to calculate another important characteristic, the
nucleon momentum distribution $n(q)$ of the deuteron using six
invariant functions $f_{i}$ $(i=1,...,6)$ instead of two ($S$- and
$D$-waves) in the nonrelativistic case. The comparison with the
$y$-scaling data shows the decisive role of the function $f_{5}$
which at $q\geq$ 500 MeV/c exceeds all other $f$-functions (as
well as the $S$- and $D$-waves) for the correct description of
$n(q)$ of the deuteron in the high-momentum region. Comparison
with other calculations using $S$- and $D$-waves corresponding to
various nucleon-nucleon potentials is made. Secondly, using clear
indications that the high-momentum components of $n(q)$ in heavier
nuclei are related to those in the deuteron, we develop an
approach within the natural orbital representation to calculate
$n(q)$ in $(A,Z)$-nuclei on the basis of the deuteron momentum
distribution. As examples, $n(q)$ in $^{4}$He, $^{12}$C and
$^{56}$Fe are calculated and good agreement with the $y$-scaling
data is obtained.
\end{abstract}

\section{Introduction}
The most recent experimental data on the deuteron structure
functions and tensor polarizations obtained at Thomas Jefferson
Laboratory have been reported in
\cite{Abo1,Abo2,Abo3,Ale99,Fur98,Bei99} (and references therein).
The new data on tensor polarization observables measured in
elastic electron-deuteron scattering for the squared momentum
transfer between 0.66 and 1.7~(GeV/c)$^{2}$ \cite{Abo1} make it
possible to determine quite precisely the deuteron charge form
factors for momenta comparable with the deuteron mass. Thus these
data are important to probe the relativistic dynamics inside the
deuteron. It was shown in \cite{Abo1} that the empirical results
for the tensor polarization $t_{20}$ and both monopole ($G_{C}$)
and quadrupole ($G_{Q}$) charge form factors are in very good
agreement with the calculations within two relativistic and
covariant models: i) a model developed in the framework of the
explicitly covariant version of light-front dynamics (LFD) (e.g.
\cite{Car99,Kar88,Car98,Car95}) and, ii) a model using a
three-dimensional reduction of the Bethe-Salpeter equation
\cite{Phi1}. At the same time the non-relativistic impulse
approximation approach \cite{Wir95} (even with inclusion of meson
exchange currents and relativistic corrections) cannot explain,
for instance, the behaviour of the monopole charge form factor in
the region of the first node and the secondary maximum shown for
the first time in \cite{Abo1}.

It is well known \cite{Jam85,Ant93,Ant93'} that one of the main
feature of the realistic nuclear models (beyond the mean-field
approximation) is that they have to describe {\it{simultaneously}}
both important ground state characteristics, the nucleon momentum
and density distributions of the system. This is not the case in
the Hartree-Fock approximation where the dynamical short-range and
tensor nucleon-nucleon (NN) correlations are only partly
incorporated. The main characteristic feature of the data on the
nucleon momentum distribution $n(q)$ extracted by the nuclear
$y$-scaling analysis \cite{Day90,Cio87,Cio90,Cio96} (we shall call
them all along this paper $y$-scaling data (YSD)) is the existence
of high-momentum components for momenta $q >2$ fm$^{-1}$ due to
the short-range and tensor NN correlations. This has been shown
also for particular nuclei such as $^{2}$H, $^{3,4}$He, $^{12}$C,
$^{16}$O, $^{40}$Ca and others and in nuclear matter within
various theoretical correlation methods (see e.g.
\cite{Ama76,Zab78,Ord80,Boh80,Ant80,Del83,Cio84,Fan84,Aka84,Fly84,Tra85,Ben86,Sch86,Jam86,Mor88,Str90,Pie90,Bal90,Ant94,Sto93,Mut95}
and reviews in \cite{Ant93,Ant93'}). In addition, both
experimental and theoretical analyses confirm the conclusion that
the high-momentum behaviour of the nucleon momentum distribution
($n(q)$/A at $q >2$ fm$^{-1}$) is similar for nuclei with mass
numbers $A$=2, 3, 4, 12 and 56 and for nuclear matter.

The first aim of the present work is to apply the LFD method to
calculations of the momentum distribution in the deuteron. Thus
this study is an {\it important complementary test} of the
possibility of the LFD {\it to describe simultaneously} both
deuteron charge form factors (that has been shown in \cite{Abo1})
and momentum distribution. In this line the results of
calculations using wave functions of existing NN potentials are
also given and compared with the LFD results and the YSD.

The second aim of the work is to develop a method which would
allow us to calculate the nucleon momentum distribution for a wide
range of $(A,Z)$-nuclei on the basis of the knowledge of $n(q)$ in
the deuteron and especially of its high-momentum components. The
latter are important because in various processes with high
transferred momenta only relativistic nucleons can take part
\cite{Kar88}. Such analyses require the knowledge of the wave
function (WF) at momenta with magnitude of the nucleon mass. We
make an attempt to relate the relativistic effects in $n(q)$ in
the deuteron with the explanation of the high-momentum components
of $n(q)$ in heavier nuclei which are important for calculations
of the cross sections of various nuclear reactions. This question
is discussed together with the effects of short-range NN
correlations on the nucleon momentum distributions.

A brief review of the LFD method is given in Sect. II. The results
of the calculations of $n(q)$ in deuteron within the LFD and the
comparison with calculations using wave functions corresponding to
existing NN potentials and the YSD are given in the same Section.
In Sect. III a method to calculate the nucleon momentum
distribution in $(A,Z)$-nuclei is developed using the natural
orbital representation and the results on the high-momentum
components of the nucleon momentum distribution in the deuteron.
Sect. IV contains the conclusions of the present work.

\section{Nucleon Momentum Distribution in the Deuteron Within the LFD Method}

The results of the experimental and theoretical studies of the
structure and form factors of the deuteron have been presented in
the most recent review \cite{Gar01}. In it special attention has
been paid to the different theoretical models for investigating
the deuteron electromagnetic form factors and nucleon momentum
distribution. Among them, the LFD method has been discussed as one
of the promising approaches. We start this Section with a brief
description of the LFD (e.g.
\cite{Car99,Kar88,Car98,Car95,Kar81,Kar92,Kar94}) which is applied
to the analyses of relativistic bound systems at high relative
momenta. The LFD is a self-consistent method in which the
relativistic wave functions are the Fock column components of the
state vector defined on an arbitrary light-front (LF) hypersurface
given by invariant equation ${\omega}x=0$. The four-vector
${\omega}=({\omega}_{0},\vec{{\omega}})$, ${\omega}^{2}=0$,
determines the position of the LF surface. The state vector
satisfies a dynamical equation following from rotations of the
light front. In LFD the transformations of the reference system
and of the light front are independent of each other. The
dynamical dependence of the state vector on the LF position is
parametrized by the explicit dependence of the wave functions on
the four-vector ${\omega}$. The dependence on ${\omega}$, however,
is not a property of the observable amplitude. The form factors
depend on four-momentum transfer squared only and do not depend on
${\omega}$, though they are related to the ${\omega}$-dependent
wave functions \cite{Car95}. The relativistic deuteron WF on the
LF $\Psi(\vec{q},\vec{n})$ depends on two vector variables: i) the
relative momentum $\vec{q}$ and ii) on a unit vector $\vec{n}$
along $\vec{\omega}$. Due to this, the WF is determined by six
invariant functions instead of two ($S$- and $D$-waves) in the
non-relativistic case. Each one of these functions depends on two
scalar variables $q$ and $z=cos(\widehat{\vec{q},\vec{n}})$. In
LFD these six functions are calculated within the relativistic
one-boson-exchange model. The kernel in the dynamical equation for
$\Psi(\vec{q},\vec{n})$ has been calculated \cite{Car95} using
Lagrangians of interaction of nucleon with pseudoscalar
$(\pi,\eta)$-, scalar $(\sigma,\delta)$- and vector
$(\rho,\omega)$ mesons using the parameters of the Bonn model
\cite{Mac87}. Thus in the LFD calculations the deuteron structure
is determined by the relativistic nucleon-meson dynamics,
supposing that the nucleons in the deuteron interact by exchanging
relativistic mesons \cite{Car99}, without using potential
approximations.

The relations of the covariant LFD method to other relativistic
approaches, such as the standard LFD \cite{Dir49}, the
Bethe-Salpeter formalism \cite{Sal51} and its three-dimensional
reductions, the quasi-potential equations \cite{Log66} and the
Gross equation method \cite{Gro82} are given in \cite{Car98}. For
instance, it is shown in \cite{Bon99,Car98} that after the
projection of the Bethe-Salpeter amplitude on the light front, the
six components of the LFD deuteron WF are expressed through
integrals over the eight components of the deuteron Bethe-Salpeter
amplitude. Provided the NN interaction is the same, these
approaches incorporate by different methods the same relativistic
dynamics. The WF's in LFD are the direct relativistic
generalization of the non-relativistic ones in the sense that they
are still the probability amplitudes. Therefore they can be used
in the relativistic nuclear physics (e.g. \cite{Car95}). Another
advantage of LFD is that due to the explicit relativistic
covariance of the method, the form factors do not depend on the
system of reference and on the direction of the $z$-axis. One
should emphasize that the LFD WF's have been successfully used
also in the hadron physics \cite{Bro98}.

As mentioned in the Introduction, the LFD results give a good
description of  the data \cite{Gar01} for the tensor polarization
$t_{20}$ and for the structure function $A(Q^{2})$ (up to
$Q^{2}$$\simeq 3$ (GeV/c)$^{2}$). Concerning the structure
function $B(Q^{2})$ its minimum occurs below the position
indicated by the data. The latter is related \cite{Car99,Gar01} to
the insufficient accuracy of the perturbative approach in
calculating the components of the LFD wave functions. We would
like to emphasize, however, that the LFD method describes very
well the charge and quadrupole form factors which are expressed by
the structure functions A(Q$^{2}$), B(Q$^{2}$) and the tensor
polarization $t_{20}$.

In this Section we firstly calculate the nucleon momentum
distribution in deuteron $n(q)$ within the LFD using the WF
$(\Psi(\vec{q},\vec{n}))$ which is normalized generalizing the
non-relativistic normalization condition \cite{Car95}:
\begin{equation}\label{norm1}
\frac{1}{3}\frac{m}{(2\pi)^{3}}\int{\Psi^{2}(\vec{q},\vec{n})}\frac{d^{3}\vec{q}}{\varepsilon(\vec{q})}=
\frac{m}{(2\pi)^{3}}\int{F(\vec{q}^{2},\vec{n}\cdot\vec{q})}\frac{d^{3}\vec{q}}{\varepsilon(\vec{q})}=1\,,
\end{equation}
where $F$ is expressed by the six scalar functions $f_{1-6}$
depending on the scalar variables $\vec{q}^{2}$ and
$\vec{n}\cdot\vec{q}$:
\begin{equation}\label{F}
F=f_{1}^{2}+f_{2}^{2}+f_{3}^{2}+(3+z^{2})f_{4}^{2}+(1-z^{2})(f_{5}^{2}+f_{6}^{2})+(3z^{2}-1)f_{2}f_{3}+4zf_{4}(f_{2}+f_{3})\,.
\end{equation}
In (\ref{norm1})
${\varepsilon(\vec{q})}=\sqrt{{\vec{q}}^{2}+m^{2}}$, where $m$ is
the nucleon mass. As shown in \cite{Car95}, in the
non-relativistic limit ($q<<m$) the functions $f_{3-6}$ become
negligible, $f_{1,2}$ do not depend on $z$ and turn into $S$- and
$D$-waves (${f_{1}}{\approx}{u_{S}}$, $f_{2}{\approx}-u_{D}$) and
the WF $\Psi(\vec{q},\vec{n})$ becomes the usual non-relativistic
wave function. One of the most important properties of the
functions $f_{1-6}$ found in \cite{Car95} is that for $q\geq$
2$\div$2.5 fm$^{-1}$ the component $f_{5}$ (being related mainly
to $\pi$-exchange) exceeds sufficiently the $S$- and $D$-waves.
This fact is very important in the calculations of $n(q)$ in
deuteron as it will be shown below.

In \cite{Car95} a Legendre expansion of $f_{1-6}(q,z)$ relative
to $z$ has been used:
\begin{equation}\label{f}
f_{i}(q,z)=\sum_{l}(2l+1)f_{i}^{l}(q)P_{l}(z)
\end{equation}
($l$=0,2 for $i$=1,2,3,5 and $l$=1,3 for $i$=4,6) with given
values of the coefficients $f_{i}^{l}(q)$.

We calculate in our work the angle-averaged nucleon momentum
distribution in the deuteron defined as:
\begin{equation}\label{n}
n(q)=C(q)\overline{F(q)},
\end{equation}
where
\begin{equation}\label{C}
C(q)=\frac{m}{(2\pi)^{3}\varepsilon(q)}
\end{equation}
and $\overline{F(q)}$ is the angle-averaged function
$F(q,z=\cos\theta)$:
\begin{equation}\label{F-sr}
\overline{F(q)}=\frac{1}{4\pi}\int{F(q,{\cos}{\theta})d{\Omega}}.
\end{equation}
In accordance with Eq. (\ref{norm1}) the normalization of $n(q)$
is given by:
\begin{equation}\label{norm2}
\int{n(\vec{q})d^{3}{\vec{q}}}=1\,.
\end{equation}

The LFD calculations have shown that, as expected, the most
important contributions to the total $n(q)$ give terms related to
the $f_{1}$, $f_{2}$ and $f_{5}$ functions
\begin{equation}\label{n1}
n(q){\simeq}n_{1}(q)+n_{2}(q)+n_{5}(q),
\end{equation}
where
\begin{equation}\label{ndef}
n_{1}(q)=C(q)\overline{f_{1}^{2}(q)},\quad
n_{2}(q)=C(q)\overline{f_{2}^{2}(q)},\quad
n_{5}(q)=C(q)\overline{(1-z^{2})f_{5}^{2}(q)}.
\end{equation}
Here we would like to note that the use of the functions $f_{1}$,
$f_{2}$ and $f_{5}$ averaged over
$z=cos(\widehat{\vec{q},\vec{n}})$ can be justified by their
smooth (almost constant) behaviour as functions of $z$ at
different values of $q$ (see Figs. 10 and 11 of \cite{Car95}). The
contributions of $n_{1}$, $n_{2}$, $n_{12}=n_{1}+n_{2}$ and
$n_{5}$ are compared in Fig. 1. It can be seen that, while the
functions $f_{1}$ and $f_{2}$ give a good description of the YSD
of $n(q)$ for $q<2$ fm$^{-1}$ (like the $S$- and $D$-wave
functions in the non-relativistic case), it is impossible to
explain the high-momentum components of $n(q)$ at $q>2$ fm$^{-1}$
without the contribution of the function $f_{5}$. We note that the
deviation of the total $n(q)$ from the sum $n_{12}=n_{1}+n_{2}$
starts at $q$ around 1.8 fm$^{-1}$. All this shows the important
role of NN interactions which incorporate exchange of relativistic
mesons in the case of the deuteron. We emphasize the
{\it{consistency of the LFD method obtained in the simultaneous
description of both deuteron momentum distribution (made in this
work) and of charge form factors}} \cite{Abo1}. Here we would like
to discuss this point. Considering the LFD results on the nucleon
momentum distribution in the deuteron we do not imply that the LFD
method and the contribution of the $f_{5}$ function are the only
way to understand the high-momentum components of $n(q)$. In the
following of this Section we will pay attention to the results of
other approaches considering, however, this question in close
connection to the possibility of simultaneous description of both
momentum distribution and form factors. It is known, for instance,
that the deuteron wave function related to the Argonne v$_{18}$
potential for the NN interaction \cite{Wir95} gives a realistic
description of $n(q)$ in the deuteron \cite{For96}. However, as
can be seen from Figs. \ref{fig2} and \ref{fig4} of Ref.
\cite{Abo1}, the recent non-relativistic impulse approximation
calculations using the Argonne v$_{18}$ potential and even those
including meson exchange currents and relativistic corrections do
not give a good agreement with the new data of \cite{Abo1} on the
tensor polarization $t_{20}$ and the charge form factors $G_{C}$
and $G_{Q}$.

We calculate in the present work also the angle-averaged nucleon
momentum distribution in the deuteron using $S$- and $D$-wave
functions $\Psi_{S}(q)$ and $\Psi_{D}(q)$ corresponding to various
NN potentials, such as the charge-dependent Bonn potential
\cite{Mac01}, the Argonne v$_{18}$ \cite{Wir95}, the Nijmegen - I
,- II and - Reid 93 \cite{Sto94} and Paris 1980 \cite{Lac80} in
the expression:
\begin{equation}\label{n10}
n(q)=\frac{1}{4\pi}[\Psi_{S}^{2}(q)+\Psi_{D}^{2}(q)]\equiv
n_{S}(q)+n_{D}(q)
\end{equation}
with
\begin{equation}\label{n11}
\int{{n(q)}d^{3}\vec{q}}=1\,.
\end{equation}
In Fig. \ref{fig2} the result for $n(q)$ using the
charge-dependent Bonn potential \cite{Mac01} is given and compared
with the YSD. As can be seen, the $D$-component of $n(q)$ is
important but even its inclusion does not give a very good
agreement with the data for $q\geq 2$ fm$^{-1}$. In the next Fig.
\ref{fig3} we present the LFD result for $n(q)$ compared with the
calculations using the WF's corresponding to Nijmegen-I,-II, -Reid
93, Argonne v$_{18}$ and Paris 1980 NN potentials and with the
$y$-scaling data. We would like to note that: i) the results of
the calculations using the NN potentials, such as Nijmegen-II,
-Reid 93, Argonne v$_{18}$ and Paris 1980 (shown in Fig.
\ref{fig3}) are in better agreement with the YSD than those using
the charge-dependent Bonn potential (Fig. \ref{fig2}). This might
be related to the fact that these potentials describe NN phase
shifts up to larger energies (e.g. the Nijmegen-II potential gives
reasonable $pp$ phase shifts up to 1.2 GeV, while the
charge-dependent one-boson exchange Bonn potential fits the
phase-shift data below 350 MeV); ii) It can be seen from Fig.
\ref{fig3} that there are small differences between the curves
corresponding to different NN potentials for $q\leq $ 3 fm$^{-1}$
(which give a good description of the YSD and almost coincide with
the LFD result) and larger ones for $q>3$ fm$^{-1}$. Large
differences take place, however, between all of them and the LFD
result for $q>3$ fm$^{-1}$. An important conclusion of this fact
is that data at higher momentum transfers are needed (and probably
a more refined analysis of the currents used for the coupling of
the electron with the nucleons) in order to distinguish the
properties of the covariant LFD method from the potential
approaches. The momentum region of $q$ just larger than 3
fm$^{-1}$ will be decisive in this sense. The main advantage of
the LFD method is, however, as mentioned already, that it
describes simultaneously the deuteron charge form factors as well,
which is not the case with e.g. the Argonne v$_{18}$ NN potential
method (see Fig. 4 in \cite{Abo1}).

\section{Nucleon Momentum Distributions in $^{4}$H\lowercase{e},
$^{12}$C and $^{56}$F\lowercase{e}}

As mentioned in the Introduction, realistic many-body calculations
(e.g.
\cite{Ama76,Zab78,Ord80,Boh80,Ant80,Del83,Cio84,Fan84,Aka84,Fly84,Tra85,Ben86,Sch86,Jam86,Mor88,Str90,Pie90,Bal90,Ant94,Sto93,Mut95,Ant93,Ant93'})
have shown the existence of high-momentum components in the
nucleon momentum distributions for studied nuclei at $q\geq 2$
fm$^{-1}$, due to presence of short-range and tensor correlations.
This conclusion agrees well with the $y$-scaling data
\cite{Day90,Cio87,Cio90,Cio96}. Secondly, it has been shown that
all nuclear momentum distributions $n_{A}(q)$ for $(A,Z)$-nuclei
($A>2$) at these high momenta are simply rescaled versions of the
nucleon momentum distribution in the deuteron $n(q)$ \cite{Far99}:
\begin{equation}\label{na}
n_{A}(q){\cong}\alpha_{A}n(q),
\end{equation}
where $\alpha_{A}$ is a constant.

Since the magnitude of the high-momentum tail of $n_{A}(q)$ is
proportional to the mass number, this effect is associated with
the nuclear interior rather than with the nuclear surface. In the
present paper we develop a model to calculate $n_{A}(q)$ for
nuclei with $A>2$ from that of the deuteron. The method has
similarities to that suggested in \cite{Gai95} on the basis of
$^4$He. However, in the present work we incorporate the analysis
performed in Sect. II for the deuteron as the simplest two-nucleon
bound system to the consideration of $n_{A}(q)$ for heavier
nuclei, namely because the high-momentum components are related to
the specific features of the two-nucleon interaction at short
distance (around the repulsive core). The latter determine the
corresponding behaviour of the nucleons in the central part of the
nucleus, where the density is higher. There the two nucleons could
be closer to one another and the short-range correlations could be
operative. In this sense, we consider the present method as a next
step and improvement on that of \cite{Gai95}.

In this Section we suggest a practical method to calculate the
nucleon momentum distribution for the general case of
$(A,Z)$-nuclei ($A>2$), while as we mentioned already, the
correlation methods consider only particular nuclei (e.g.
$^{3,4}$He, $^{12}$C, $^{16}$O, $^{40}$Ca and nuclear matter) and
due to their complexity the applications to other nuclei is often
a very difficult task.

We use the natural orbital representation (NOR)
\cite{Low55,Ant93,Ant93'} of the one-body density matrix. In our
case we consider its diagonal elements in the momentum space,
which represent the nucleon momentum distribution $n_{A}(q)$. The
NOR allows us to use the transparency of the single-particle
picture even being beyond the mean-field approximation and to
account for the correlation effects in a proper way. Using the
common theoretical ground of NOR, the method makes it possible to
combine the mean-field predictions for $n_{A}(q)$ (which are
expected to be realistic at $q<2$ fm$^{-1}$) with that part of the
momentum distribution which includes correlation and relativistic
effects.

In the natural orbital representation the momentum distribution
in a nucleus with $A$ nucleons can be written in the form:
\begin{equation}\label{na1}
n_{A}(q)=N_{A}[n^{h}_{A}(q)+n^{p}_{A}(q)],
\end{equation}
where $N_{A}$ is the normalization constant and
\begin{equation}\label{nah}
n^{h}_{A}(q)=\frac{1}{A}\sum_{nlj}^{F.L.}{2(2j+1)
\lambda_{nlj}C(q){\mid}R_{nlj}(q){\mid}^{2}}
\end{equation}
\begin{equation}\label{nap}
n^{p}_{A}(q)=\frac{1}{A}\sum_{F.L.}^{\infty}{2(2j+1)
\lambda_{nlj}C(q){\mid}R_{nlj}(q){\mid}^{2}}
\end{equation}
are the hole- and particle-state contributions, respectively. In
(\ref{nah}) and (\ref{nap}) F.L. denotes the Fermi level, $C(q)$
is given by (\ref{C}), $\lambda_{nlj}$ are the natural occupation
numbers ($0\leq\lambda_{nlj}\leq1$) for a state with quantum
numbers $nlj$ and $R_{nlj}(q)$ are the natural orbitals. We call
hole-state natural orbitals those natural orbitals for which the
numbers $\lambda_{nlj}$ are significantly larger than the
remaining ones, called particle-state natural orbitals
\cite{Lew88}. As shown in \cite{Sto93}, the high-momentum
components of the total $n_{A}(q)$ caused by short-range
correlations are almost completely determined by the contributions
of the particle-state natural orbitals. This fact, together with
the approximate equality of the high-momentum tails of $n(q)/A$
for all nuclei allows us to make the main assumption of the method
that the particle-state contributions to $n_{A}(q)$ are almost
equal for all nuclei. The deuteron $n(q)$ from (\ref{n1}) can be
written in a form similar to (\ref{na1}):
\begin{equation}\label{na2}
n(q)=N_{d}[n^{h}_{d}(q)+n^{p}_{d}(q)]
\end{equation}
with
\begin{equation}\label{na3}
n_{d}^{h}(q)=\frac{1}{2}[n_{1}(q)+n_{2}(q)],
\end{equation}
\begin{equation}\label{na3'}
n_{d}^{p}(q)=\frac{1}{2}n_{5}(q),
\end{equation}
and $N_{d}=2$. The assumed equality of the particle-state
contributions for all nuclei implies the equality of the
particle-state contribution $n_{A}^{p}(q)$ for $(A,Z)$-nuclei from
(\ref{nap}) with that to the deuteron momentum distribution
$n_{d}^{p}(q)$ from (\ref{na3'})
\begin{equation}\label{na3''}
n_{A}^{p}(q){\simeq}n^{p}_{d}(q).
\end{equation}

The right-hand side of (\ref{na3''}) can be taken from different
sources. This could be the YSD or results of calculations within
given theoretical correlation models. Having the results from
Sect. II we take $n_{d}^{p}(q)$ to be related to $n_{5}(q)$
(\ref{na3'}) because within the LFD namely this function is
responsible for the high-momentum tail of $n(q)$ for deuteron:
\begin{equation}\label{nd}
n_{A}^{p}(q){\cong}\frac{1}{2}n_{5}(q)\,.
\end{equation}
Then the resulting momentum distribution for $(A,Z)$-nucleus
normalized to unity has the form:
\begin{equation}\label{naq1}
n_{A}(q)=N_{A}[\sum_{nlj}^{F.L.}{2(2j+1)
\lambda_{nlj}C(q){\mid}R_{nlj}(q){\mid}^{2}}+\frac{A}{2}n_{5}(q)]\,,
\end{equation}
where
\begin{equation}\label{normnaq1}
N_{A}=\{4{\pi}{\int}dq{\cdot}q^{2}[\sum_{nlj}^{F.L.}{2(2j+1)
\lambda_{nlj}C(q){\mid}R_{nlj}(q){\mid}^{2}}+\frac{A}{2}n_{5}(q)]\}^{-1}\,.
\end{equation}
It is known from \cite{Sto93} that the hole-state natural orbitals
are almost unaffected by the short-range correlations and,
therefore, the functions $R_{nlj}(q)$ in (\ref{naq1}) and
(\ref{normnaq1}) can be replaced by single-particle wave functions
from the mean-field approximation (shell-model or Hartree-Fock
WF). In our calculations we use Woods-Saxon s.p. wave functions
for protons and neutrons. The hole-state occupation numbers
$\lambda_{nlj}$ are close to unity and we set them equal to unity
with good approximation. These properties lead to a similarity of
our model to that suggested for calculations of the spectral
function \cite{Ben94}.

The calculations of $n_{A}(q)$ for $^{4}$He,$^{12}$C and $^{56}$Fe
are given in Figs. \ref{fig4}-\ref{fig6} and are compared with the
$y$-scaling data from \cite{Cio90}. We note that, as expected from
(\ref{na}) the value of $\alpha_{A}$ must be almost constant. In
our case $\alpha_{A}=N_{A}\cdot{A}/{2}$ and its numerical values
are 8.335, 8.352 and 8.372 for $^{4}$He, $^{12}$C and $^{56}$Fe,
respectively.

The dashed line in Figs. \ref{fig4}-\ref{fig6} gives the
calculated normalized hole-state contribution (the mean-field
component) to $n_{A}(q)$ only, i.e. without the contribution of
$n_{5}(q)$ in Eq. (\ref{naq1}). The comparison between the total
$n_{A}(q)$, its hole-state contribution and the $y$-scaling data
clearly shows the role of $n_{5}(q)$ contribution. It can be seen
that the difference between the results with $n_{5}(q)$ and
without it starts for all three nuclei at $q\simeq 1.5$ fm$^{-1}$
and increase rapidly at $q>1.5$ fm$^{-1}$. This shows the role of
the high-momentum components of the deuteron momentum distribution
$n_{5}$ which can be considered as an "ingredient" of the
high-momentum components of the momentum distribution of
$(A,Z)$-nucleus. In our opinion, the related function $f_{5}$  of
the LFD method incorporates the main part of the short-range
features of the NN interactions which determine the correlation
effects seen in $n_{A}(q)$ when calculated within the
non-relativistic correlation methods.

The good agreement of the results on $n_{A}(q)$ with the data for
the three nuclei achieved, including also the region of the high
momenta, proves the assumption that the nucleon momentum
distribution in nuclei can be extracted on the basis of the
realistically described properties of the deuteron as the smallest
bound system of two nucleons.

\section{Conclusions}

The results of the present work can be summarized as follows:

(i) The light-front dynamics method is applied to calculate the
nucleon momentum distribution in the deuteron. A very good
description of the $y$-scaling data is obtained. It is shown that,
while the functions $f_{1}$ and $f_{2}$ describe well the YSD of
$n(q)$ for $q<2$ fm$^{-1}$, it is impossible to explain the
high-momentum components of $n(q)$ at $q>2$ fm$^{-1}$ without the
contribution of the function $f_{5}$. Thus, this study gives a
positive answer to the question about the possibilities of the LFD
method {\it to describe simultaneously} both deuteron momentum
distribution and charge form factors (the good agreement with the
latter as well as with the data for the tensor polarization
$t_{20}$ and for the structure function $A(Q^2)$ up to
$Q^{2}\simeq 3$ (GeV/c)$^{2}$ was shown in \cite{Abo1,Gar01}).

(ii) The nucleon momentum distribution in deuteron was calculated
also in this work using the $S$- and $D$-wave functions
corresponding to various NN potentials (charge-dependent Bonn,
Paris 1980, Nijmegen-I and -II, -Reid 93, Argonne v$_{18}$). We
show that the results using these potentials explain very well
(with the exception of the charge-dependent Bonn potential) all
the available data for $n(q)$ up to $q\simeq 3$ fm$^{-1}$ exactly
like the LFD method. However, as shown in \cite{Abo1}, the recent
non-relativistic impulse approximation calculations (e.g. with
Argonne v$_{18}$ potential) and even those including meson
exchange currents and relativistic corrections do not describe
well the new data \cite{Abo1} on the tensor polarization $t_{20}$
and the charge form factors $G_{C}$ and $G_{Q}$. This is the main
difference between the results of the LFD method and of the
calculations using WF's corresponding to NN potentials for the
important deutron characteristics considered. Secondly, we should
mention also that for $q>3$ fm$^{-1}$ the LFD results for $n(q)$
deviate strongly from those of the calculations using NN
potentials. Thus data at higher $q$ are needed in order to
distinguish the abilities of the LFD from those of the potential
methods. Here we would like to mention also other relativistic
methods such as the method in \cite{Phi1} (using a
three-dimensional reduction of the Bethe- Salpeter  equation) and
the method in \cite{vanor95} (based on the Gross equation
\cite{Gro82}) which could give a simultaneous description  of both
deuteron form factors and momentum distribution.

(iii) A correlation method for calculating the nucleon momentum
distribution in nuclei with $A>2$ $n_{A}(q)$ is proposed. The
method uses the natural orbital representation of the one-body
density matrix and combines the mean-field component of $n_{A}(q)$
with the correlated high-momentum components taken from the
deuteron momentum distribution. The method is based on the
well-known results of realistic many-body calculations which have
shown that, due to the short-range and tensor NN correlations, all
nuclear momentum distributions $n_{A}(q)$ at high momenta ($q>2$
fm$^{-1}$) are rescaled versions of the nucleon momentum
distribution in the deuteron $n(q)$. In the present work we
incorporate in the method the LFD results for $n(q)$. The role of
the wave function $f_{5}$ is clearly shown by the comparison of
the calculations of $n_{A}(q)$ using the hole-state (mean-field)
contribution without and with an inclusion of the $n_{5}$($f_{5}$)
contribution. In our opinion, the LFD function $f_{5}$
incorporates the main part of the short-range features of the NN
interactions which can be seen in calculations of $n_{A}(q)$
within non-relativistic correlation methods. The suggested method
gives an easy practical way for calculations of $n_{A}(q)$ for any
nucleus because only hole-state wave functions and occupation
probabilities enter its main relationships (Eqs.(\ref{naq1}) and
(\ref{normnaq1})). The method gives a good description of the
$y$-scaling data for $n_{A}(q)$ in $^{4}$He, $^{12}$C and
$^{56}$Fe including the high-momentum region, showing a clear
difference from the mean-field predictions for $q>1.5$ fm$^{-1}$.

\section{Acknowledgments}

One of the authors (A.N.A.) thanks Prof. V.A. Karmanov and Prof. J. Carbonell
for the helpful discussions. He is also grateful to the Royal Society in
London and the Bulgarian Academy of Sciences for support during
his visit to the University of Oxford. The work was partly
supported by the Bulgarian National Science Foundation (Contracts
No.$\Phi-809$ and $\Phi-905$).

\newpage

\begin{figure}
\caption{The nucleon momentum distribution in deuteron. The
contributions of $n_{1}$, $n_{2}$, $n_{12}=n_{1}+n_{2}$ and
$n_{5}$ are presented. The $y$-scaling data are from
\protect\cite{Cio90}. The normalization is:
$\int{{n(q)}d^{3}\vec{q}}=1$. \label{fig1}}
\end{figure}

\begin{figure}
\caption{The nucleon momentum distribution in deuteron (solid
line) calculated using Eqs. (\ref{n10}) and (\ref{n11}) with $S$-
and $D$-wave functions corresponding to the charge-dependent Bonn
potential \protect\cite{Mac01}. $S$- and $D$-contributions are
given by dashed and dotted line, respectively. The $y$-scaling
data are from \protect\cite{Cio90}. \label{fig2}}
\end{figure}

\begin{figure}
\caption{The nucleon momentum distribution in deuteron calculated
with the LFD (solid line) in comparison with the results presented
by numerated arrows, as follows: 1-Argonne v$_{18}$, 2-Nijmegen
Reid 93, 3-Nijmegen I, 4-Nijmegen II, 5-Paris 1980 NN potentials
and with the $y$-scaling data \protect\cite{Cio90}. \label{fig3}}
\end{figure}

\begin{figure}
\caption{The nucleon momentum distribution in $^{4}$He calculated
using Eqs. (\ref{naq1}) and (\ref{normnaq1}) (solid line). The
dotted line represents the hole-state  contribution only. The
$y$-scaling data are from \protect\cite{Cio90}. The normalization
is: $\int{{n_{A}(q)}d^{3}\vec{q}}=1$. \label{fig4}}
\end{figure}

\begin{figure}
\caption{The same as in Fig.\ref{fig4} for $^{12}$C. \label{fig5}}
\end{figure}

\begin{figure}
\caption{The same as in Fig.\ref{fig4} for $^{56}$Fe.
\label{fig6}}
\end{figure}

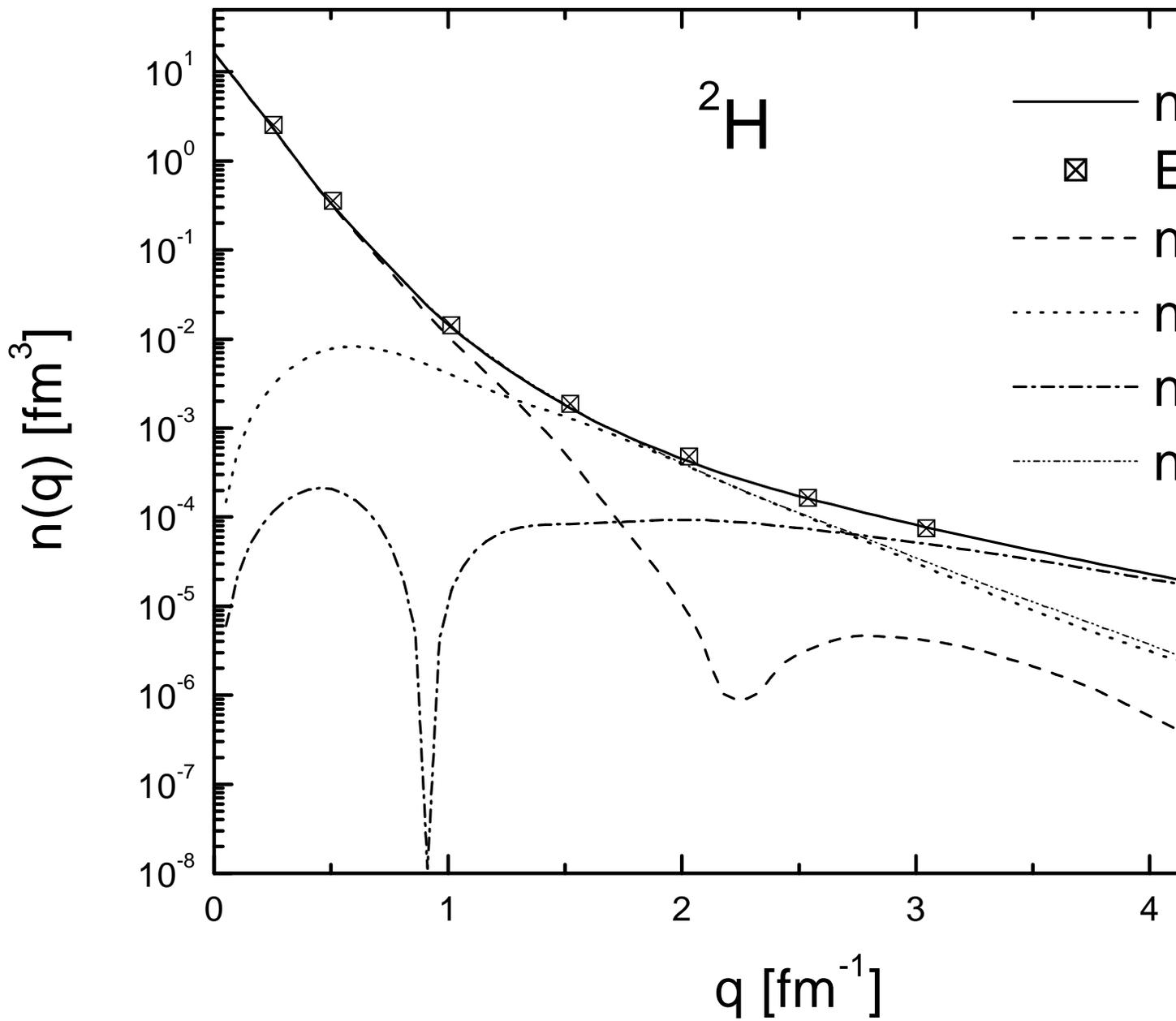
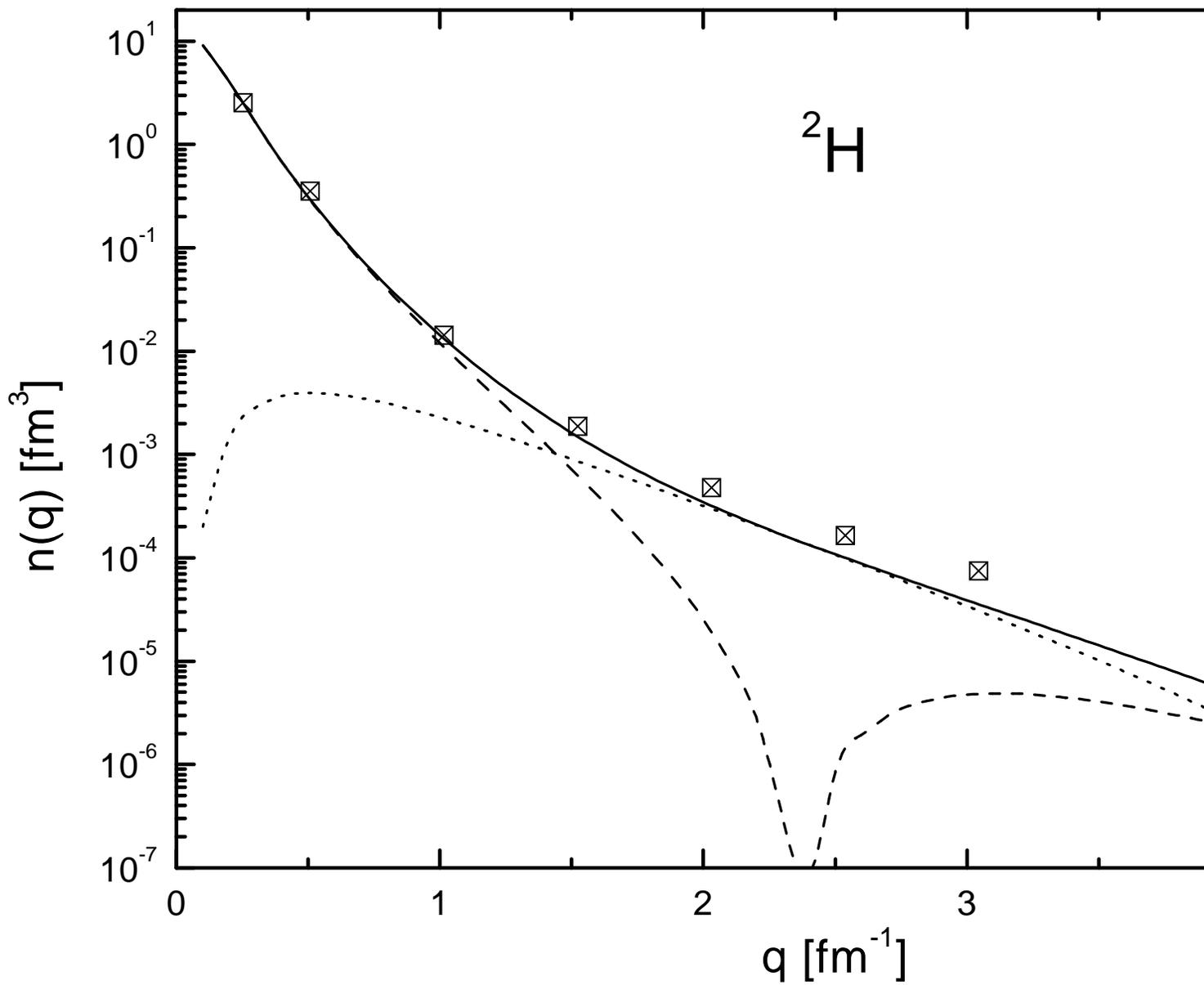
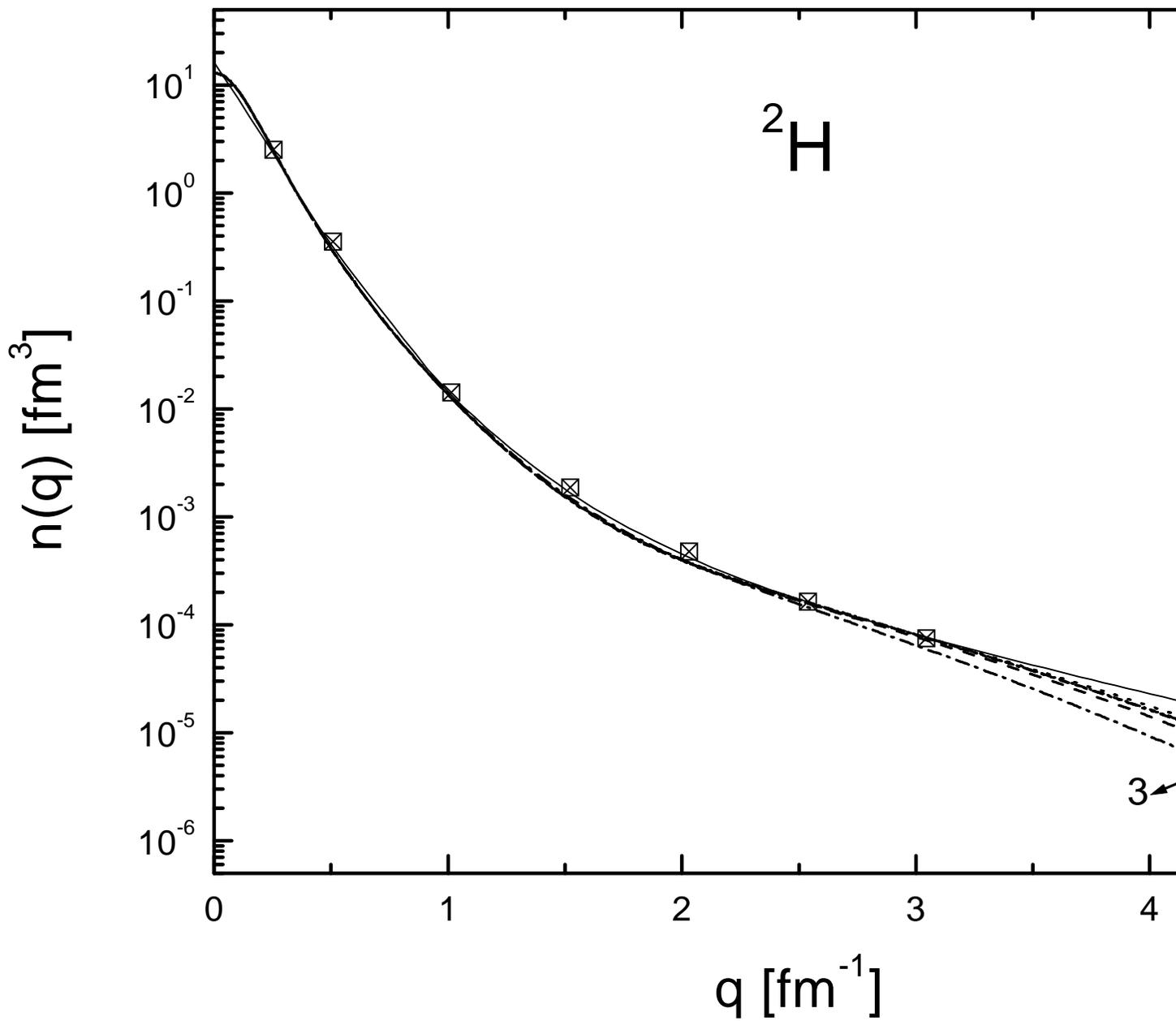
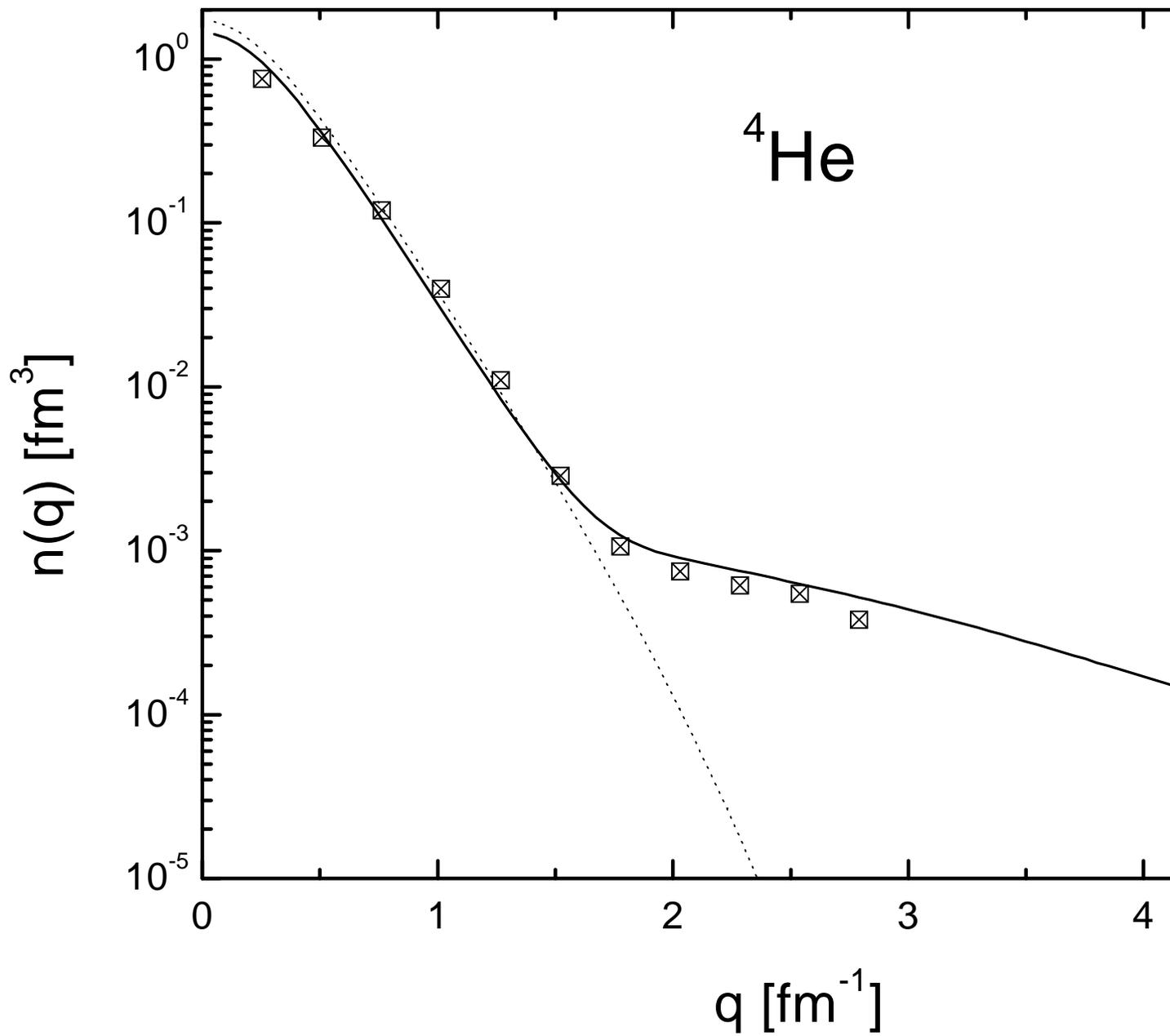
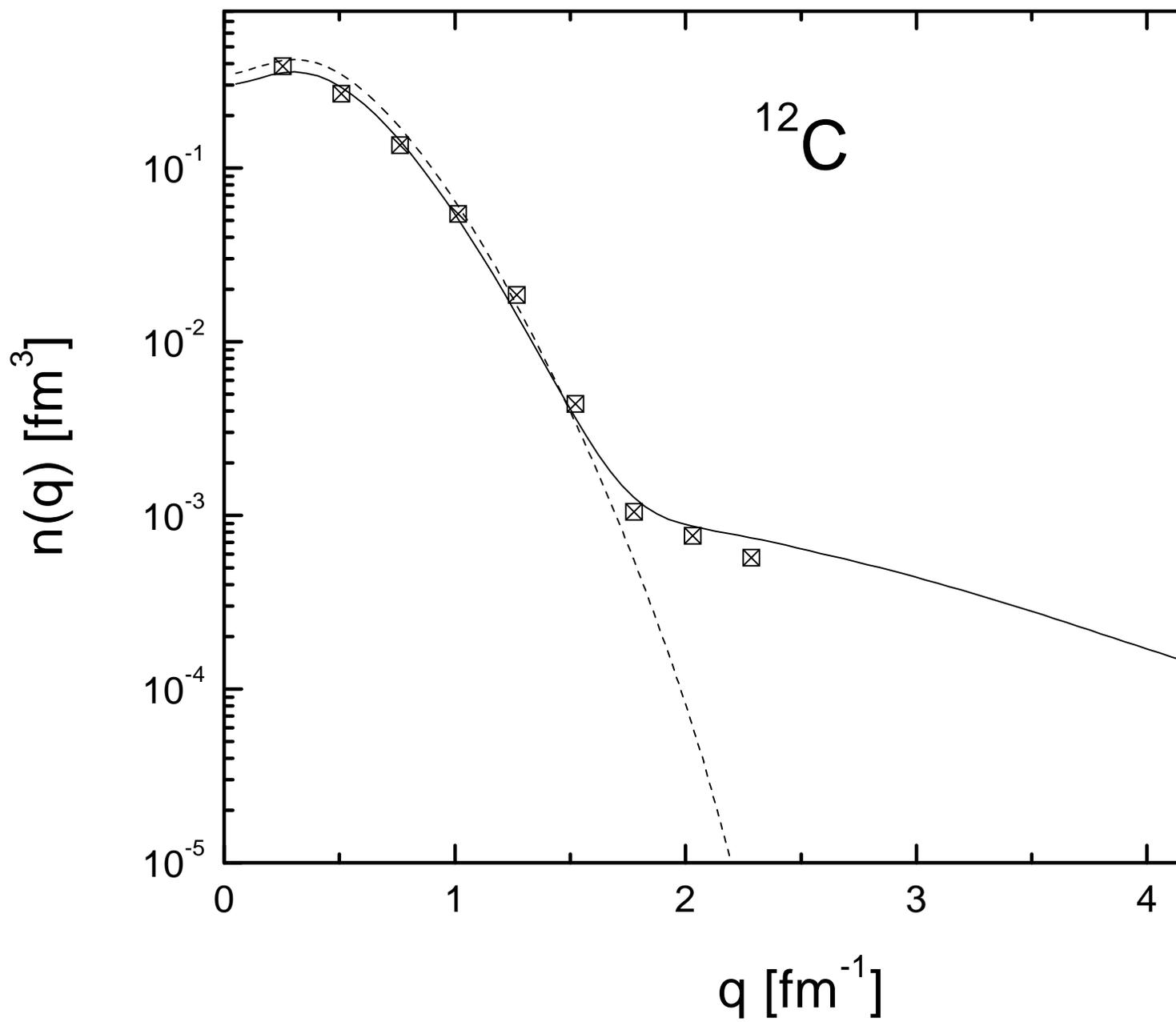
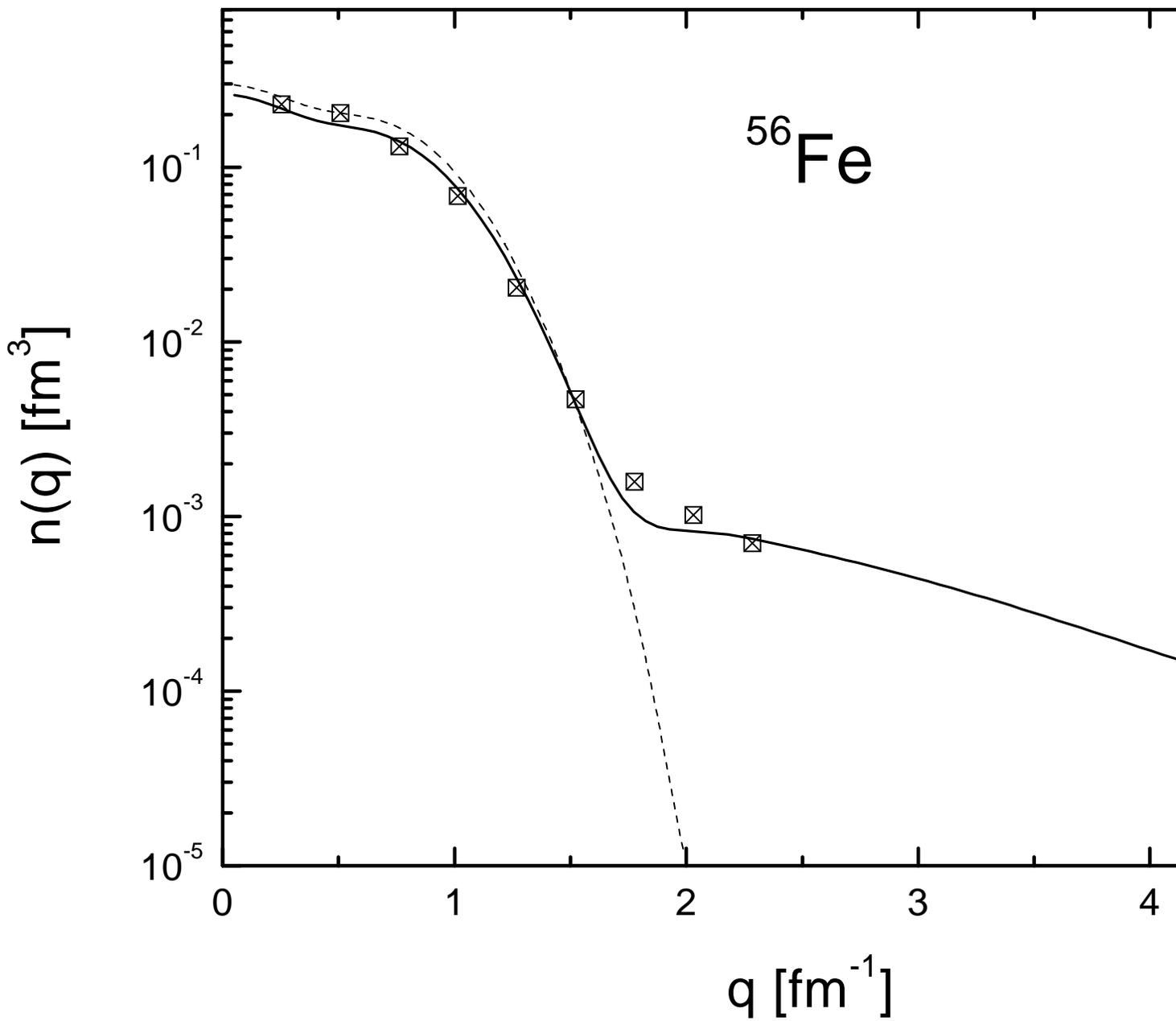
\end{document}